\documentclass[conference]{IEEEtran}
\IEEEoverridecommandlockouts
\usepackage{cite}
\usepackage{amsmath,amssymb,amsfonts}
\usepackage{algorithmic}
\usepackage{graphicx}
\usepackage{textcomp}
\usepackage{xcolor}

\usepackage{hyperref}

\usepackage{comment}
\newboolean{showcomments}
\setboolean{showcomments}{true}
\ifthenelse{\boolean{showcomments}}
{ \newcommand{\mynote}[3]{
    \fbox{\bfseries\sffamily\scriptsize#1}
    {\small$\blacktriangleright$\textsf{\emph{\color{#3}{#2}}}$\blacktriangleleft$}}}
{ \newcommand{\mynote}[3]{}}
\newcommand{\shrink}[1]{}

\def\BibTeX{{\rm B\kern-.05em{\sc i\kern-.025em b}\kern-.08em
    T\kern-.1667em\lower.7ex\hbox{E}\kern-.125emX}}

\RequirePackage{circledsteps}
\pgfkeys{/csteps/inner color=white}
\pgfkeys{/csteps/fill color=black}

\begin{document}

\title{A retrospective on DISPEED -- Leveraging heterogeneity in a drone swarm for IDS execution
\thanks{AID: Agence de l'Innovation de Défense}
}

\author{
    \IEEEauthorblockN{
        Vincent Lannurien\IEEEauthorrefmark{1}, 
        Camélia Slimani\IEEEauthorrefmark{1},
        Louis Morge-Rollet\IEEEauthorrefmark{1}, \\
        Laurent Lemarchand\IEEEauthorrefmark{2},
        David Espes\IEEEauthorrefmark{2},
        Frédéric Le Roy\IEEEauthorrefmark{1}, 
        Jalil Boukhobza\IEEEauthorrefmark{1}
    }

    \IEEEauthorblockA{
        Lab-STICC, CNRS, UMR 6285, 
        $\{$ \IEEEauthorrefmark{1}~ENSTA, Institut Polytechnique de Paris,
        \IEEEauthorrefmark{2}~Université de Bretagne Occidentale $\}$, Brest \\
        Email: $\{$vincent.lannurien, frederic.leroy, jalil.boukhobza$\}$@ensta.fr, \\
        camelia.slimani@toulouse-inp.fr, louis.morge-rollet@grenoble-inp.fr,
        laurent.lemarchand@univ-brest.fr
   }
}

\maketitle

\begin{abstract} 
    Swarms of drones are gaining more and more autonomy and efficiency during their missions. However, security threats can disrupt their missions' progression. To overcome this problem, Network Intrusion Detection Systems ((N)IDS) are promising solutions to detect malicious behavior on network traffic. However, modern NIDS rely on resource-hungry machine learning techniques, that can be difficult to deploy on a swarm of drones.
    The goal of the DISPEED project is to leverage the heterogeneity (execution platforms, memory) of the drones composing a swarm to deploy NIDS. It is decomposed in two phases: (1) a characterization phase that consists in characterizing various IDS implementations on diverse embedded platforms, and (2) an IDS implementation mapping phase that seeks to develop selection strategies to choose the most relevant NIDS depending on the context.
    On the one hand, the characterization phase allowed us to identify 36 relevant IDS implementations on three different embedded platforms: a Raspberry Pi 4B, a Jetson Xavier, and a Pynq-Z2. On the other hand, the IDS implementation mapping phase allowed us to design both standalone and distributed strategies to choose the best NIDSs to deploy depending on the context. The results of the project have led to three publications in international conferences, and one publication in a journal. 
\end{abstract}


\section{Introduction}

Unmanned Surface Vehicles (USVs) are used to carry out large-scale missions, which can involve drones with high computing power and autonomy, as well as less expensive drones with limited computing and battery capacity.

USVs operating in swarms (cooperation between USVs) can accomplish far more complex missions. However, this implies a high level of communication between drones, which can expose them to a variety of attacks. It is therefore necessary to detect attempted intrusions in good time and at low energy cost, using Intrusion Detection Systems (IDS).

Modern IDSs are mostly based on machine learning (ML) methods, where models are trained to identify abnormal and potentially malicious traffic. However, running inferences on ML models is generally costly in terms of computing and storage resources (main memory and disks), as well as consuming energy. It is therefore essential to optimize their execution so that the IDS task has the least impact on resource use, for USVs to accomplish their main tasks while ensuring a satisfactory level of safety.

One possible approach is to take advantage of the hardware heterogeneity that can exist among USVs, to choose the hardware configurations best suited both to the level of security required according to the geographical area in which the swarm is evolving, and to the resources available (computing and memory capacity) on the USV.

\section{Overview of the DISPEED project}

The aim of DISPEED is to explore possible trade-offs in terms of safety, performance and energy for executing IDS on a swarm of USVs by exploiting intra/inter-USV hardware heterogeneity. Figure~\ref{fig:overview} shows the general operation of the platform designed as part of the project:

\begin{figure*}[t]
    \centering
    \includegraphics[width=0.9\textwidth]{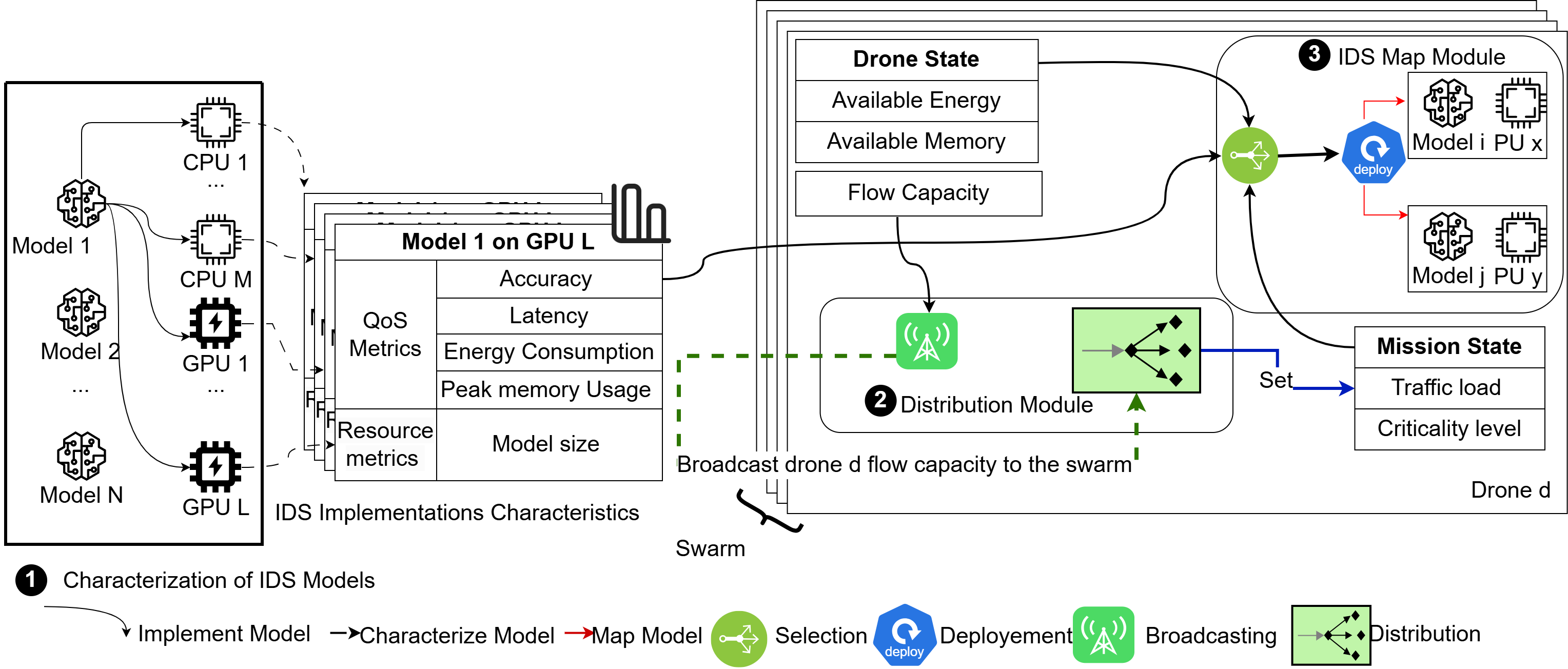}
    \caption{High-level overview of the final system considered in DISPEED.}
    \label{fig:overview}
\end{figure*}

\begin{itemize}
    \item \Circled{1} \textbf{Characterization of IDS models and execution platforms}: IDSs are based on different machine learning algorithms (e.g. random forests, DNN) and deployed on heterogeneous computing elements (e.g. CPU, GPU, FPGA). We propose an offline methodology to characterize this environment in terms of Quality of Service (QoS) -- latency, accuracy, energy consumption -- and resource metrics (memory usage, storage, etc.)~\cite{slimaniDSD23, slimani2025study}. A key finding of this study is the significant disparity in implementations characteristics, with many metrics exhibiting trade-offs (e.g., higher accuracy often comes at the cost of increased latency or energy consumption). Consequently, the optimal IDS implementation depends on the specific constraints of a given mission, motivating further investigation in studies \Circled{2} and \Circled{3};  
    \item \Circled{2} \textbf{Distribution of traffic to be analyzed within the swarm}: based on measurements from the offline phase, an online optimization strategy is implemented to decide on a distribution of the traffic to be analyzed between the drones in the swarm, so as to make the best compromise between energy and QoS~\cite{DisPEEDDistributingPacket2024}. The distribution module is decomposed in two steps: (1) drone capacity self-assessment, and (2) flow distribution. On the one hand, the drone capacity self-assessment allows each drone to estimate its processing capacity, as well as its workload, and broadcast it to the rest of the swarm. On the other hand, the flow distribution allows determining how the swarm will process the packet flows, while trying to minimize communications overhead. At the end of this phase, each drone can estimate its traffic load for the future phase; 
    \item \Circled{3} \textbf{Mapping IDSs to USVs}: the measurements from the characterization phase are used as part of a mixed offline/online optimization strategy that, depending on the state of each USV in the swarm, the characteristics of the IDS models and the state of the mission, selects the IDS model best suited to the situation and deploys it on the USV~\cite{morgeRolletIDSDEEPStrategySelectinga}. The selection process is divided in two phases: (1) the offline phase, which objective is to select implementations from the characterized set that lie on the Pareto front, while ensuring they meet the USV's storage constraints; and (2) the online phase, which consists in filtering the implementations that satisfy the live mission constraints, and choosing among them the implementation that hits the best trade-off between QoS metrics. 
\end{itemize}

The aim is to explore different models of IDS, and implement them on different hardware architectures, in order to extract data on the level of security guaranteed by the model, as well as the time and energy cost of its implementations. Given the characteristics of all the implementations, the aim is to select the best implementation during the mission, \textit{i.e.} the one that achieves the best compromise between security, performance and energy.

This heterogeneity can also be exploited to distribute the load in the swarm. The availability of each drone in the swarm can change during the course of a mission - depending on the criticality of the tasks allocated to it, but also in relation to a possible breakdown, for example. In this way, the distribution of the analysis work to be carried out on each drone can evolve during the course of a mission, so as to guarantee QoS. This flow distribution in a distributed system like a swarm of drones is not trivial, and requires rigorous analysis and optimization techniques to guarantee intrusion detection within the allotted time.








\section{Conclusion}

In DISPEED, we devised a framework for IDS deployment on a swarm of heterogeneous drones. In order to optimize the system for energy consumption while enforcing QoS under security constraints, our solution considers the system at the granularity of the swarm to optimize traffic distribution across the drones; and at the granularity of a drone to optimize IDS selection. To sum up our general approach, DISPEED leverages hardware heterogeneity across a swarm of edge devices to satisfy resource constraints as well as operational constraints during various missions. Perspectives for future work include considering the actual deployment phase. Indeed, drones are mixed-criticality systems hosting workloads that compete for shared resources. As interferences between various processes arise on such capacity-limited devices, a scheduling strategy that consider individual drones as well as the swarm as a whole might be necessary to maintain adequate levels of QoS.

\bibliographystyle{IEEEtran}
\bibliography{main}

\begin{thebibliography}{1}
\providecommand{\url}[1]{#1}
\csname url@samestyle\endcsname
\providecommand{\newblock}{\relax}
\providecommand{\bibinfo}[2]{#2}
\providecommand{\BIBentrySTDinterwordspacing}{\spaceskip=0pt\relax}
\providecommand{\BIBentryALTinterwordstretchfactor}{4}
\providecommand{\BIBentryALTinterwordspacing}{\spaceskip=\fontdimen2\font plus
\BIBentryALTinterwordstretchfactor\fontdimen3\font minus \fontdimen4\font\relax}
\providecommand{\BIBforeignlanguage}[2]{{%
\expandafter\ifx\csname l@#1\endcsname\relax
\typeout{** WARNING: IEEEtran.bst: No hyphenation pattern has been}%
\typeout{** loaded for the language `#1'. Using the pattern for}%
\typeout{** the default language instead.}%
\else
\language=\csname l@#1\endcsname
\fi
#2}}
\providecommand{\BIBdecl}{\relax}
\BIBdecl

\bibitem{slimaniDSD23}
C.~Slimani, L.~Morge-Rollet, L.~Lemarchand, F.~Le~Roy, D.~Espes, and J.~Boukhobza, ``Characterizing intrusion detection systems on heterogeneous embedded platforms,'' in \emph{2023 26th Euromicro Conference on Digital System Design (DSD)}, Durres, Albania, Sep. 2023, pp. 278--285.

\bibitem{slimani2025study}
C.~Slimani, L.~Morge-Rollet, L.~Lemarchand, D.~Espes, F.~Le~Roy, and J.~Boukhobza, ``A study on characterizing energy, latency and security for intrusion detection systems on heterogeneous embedded platforms,'' \emph{Future Generation Computer Systems}, vol. 162, p. 107473, 2025.

\bibitem{DisPEEDDistributingPacket2024}
L.~{Morge-Rollet}, C.~Slimani, L.~Lemarchand, D.~Espes, F.~Le~Roy, and J.~Boukhobza, ``{{DisPEED}}: {{Distributing Packet}} flow analysis in swarm of heterogeneous {{EmbEddeD}} platforms,'' in \emph{2025 Design, Automation \& Test in Europe Conference \& Exhibition (DATE)}, 2025.

\bibitem{morgeRolletIDSDEEPStrategySelectinga}
L.~Morge-Rollet, C.~Slimani, L.~Lemarchand, F.~L. Roy, D.~Espes, and J.~Boukhobza, ``{IDS-DEEP: a strategy for selecting the best IDS for Drones with heterogeneous EmbEdded Platforms},'' in \emph{2024 IEEE 36th International Symposium on Computer Architecture and High Performance Computing (SBAC-PAD)}.\hskip 1em plus 0.5em minus 0.4em\relax Los Alamitos, CA, USA: IEEE Computer Society, Nov. 2024, pp. 138--147.

\end{thebibliography}

\end{document}